\documentclass[aps,prb,twocolumn,groupedaddress]{revtex4}
\usepackage{amssymb,amsmath}
\usepackage{hyperref}
\usepackage[pdftex]{graphicx}

\begin{document}

\title{Spin-filtering and Disorder Induced Magnetoresistance in Carbon Nanotubes: {\it Ab Initio} Calculations}

\author{J. M. de Almeida$^1$}
\email{james.almeida@ufabc.edu.br}
\author{A. R. Rocha$^1$}
\author{Ant\^onio J. R. da Silva$^{2,3}$}
\email{jose.roque@lnls.br}
\author{A. Fazzio$^2$}
\email{fazzio@if.usp.br}
\affiliation{$^1$Centro de Ci\^encias Naturais e Humanas, Universidade Federal do ABC, Santo Andr\'e, SP, Brazil}
\affiliation{$^2$Instituto de F\'{i}sica, Universidade de S\~ao Paulo, CP 66318, 05315-970, S\~ao Paulo, SP, Brazil}
\affiliation{$^3$ Laborat\'orio Nacional de Luz S\'{\i}ncrotron, Campinas, SP, Brazil}

\begin{abstract}
Nitrogen-doped carbon nanotubes can provide reactive sites on the porphyrin-like defects. It's well known that many porphyrins have transition metal atoms, and we have explored transition metal atoms bonded to those porphyrin-like defects in N-doped carbon nanotubes. The electronic structure and transport are analized by means of a combination of density functional theory and recursive Green's functions methods. The results determined the Heme B-like defect (an iron atom bonded to four nitrogens) as the most stable and with a higher polarization current for a single defect. With randomly positioned Heme B-defects in a few hundred nanometers long nanotubes the polarization reaches near 100\% meaning an effective spin filter. A disorder induced magnetoresistance effect is also observed in those long nanotubes, values as high as 20000\% are calculated with non-magnectic eletrodes.

\end{abstract}

\maketitle

Since their discovery by Iijima in 1991, carbon nanotubes\cite{iijima,iijima2} (CNT) have become the subject of intense research due to their potential for applications,\cite{book_dresselhaus}  such as in novel electronic devices.\cite{Bachtold09112001, ISI:000073497500042} Furthermore, in a seminal paper by Tsukagoshi \textit{et al} CNTs entered the realm of spintronics, whereby one envision the possibility of using the electron spin, instead of its charge, as information carrier.\cite{tsukagoshinature_cnt} In that work the authors demonstrated that the spin coherence length of polarized electrons injected onto CNTs is larger than 300 nm. Thus, carbon nanotube devices could be used to manipulate spins in a coherent manner.

The prototypical spintronics device uses spin-polarized electrons, which are injected from a source into an unpolarized region and analyzed by a polarized drain. Within this arrangement the so-called giant magnetoresistance effect (GMR)\cite{baibich_gmr,binasch_gmr} manifests itself by altering - via an external magnetic field - the relative orientations of the magnetic moments of the electrodes. From a  practical point of view this setup usually involves sandwiching different materials. An alternative to this has been given by Kirwan et al. whereby initially unpolarized electrons are scattered by magnetic impurities adsorbed on the surface of a segment of a carbon nanotube. This way, both the electrodes as well as the device itself are made of the same material.

An alternative to this has been given by Kirwan {\it et al.}~\cite{ISI:000268429000030} whereby initially unpolarized electrons are scattered by magnetic impurities adsorbed on the surface of a segment of a carbon nanotube.\cite{ISI:000268429000030} This way, both the electrodes as well as the device itself are made of the same material, thus avoiding issues related to surface matching at the interface consequently hindering spurious scattering. 

\begin{figure}[ht]
\includegraphics[width=0.50\textwidth]{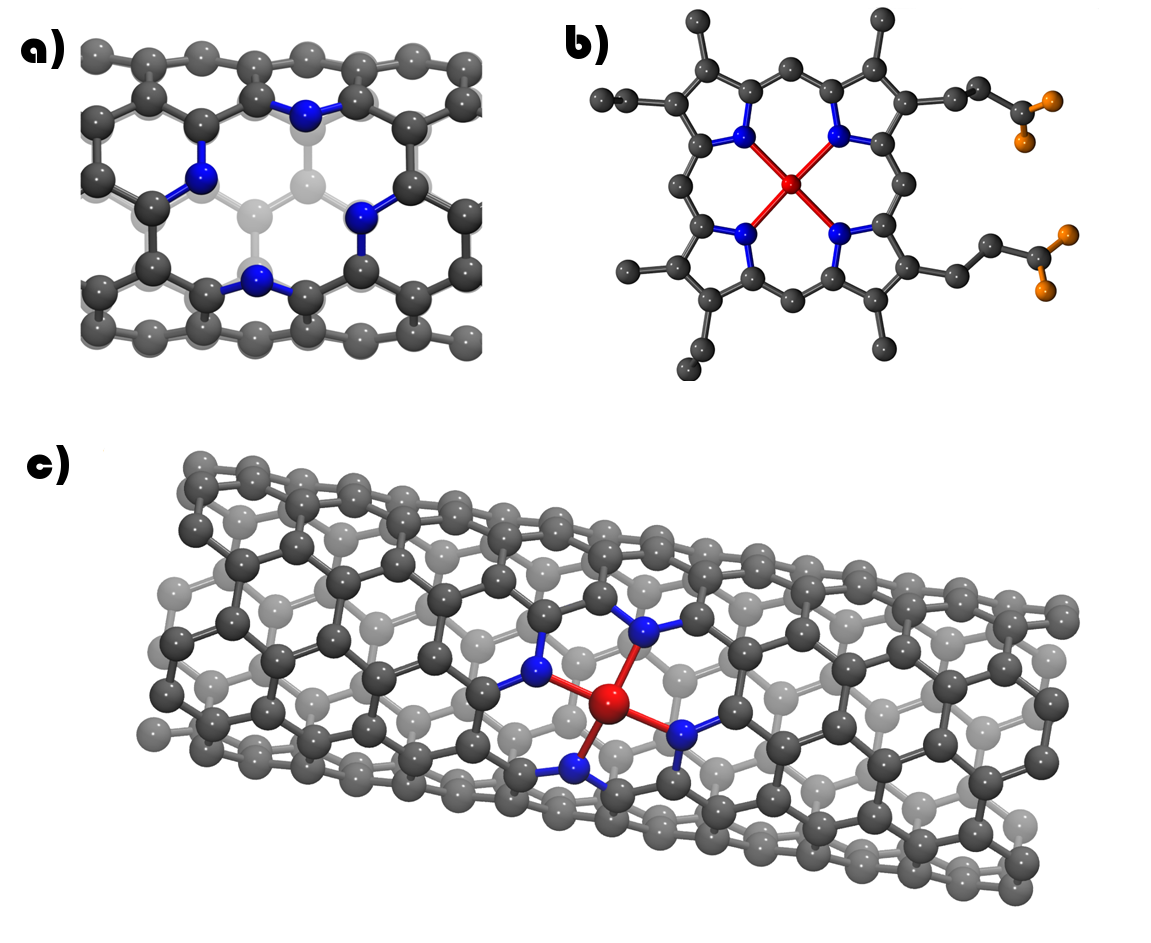}
\caption{Images of the a) Porphyrin-like defect in a carbon nanotube. b) Heme-B molecule, with an iron atom bonded to the four nitrogen atoms. c) Segment of a nanotube containing an Heme B-like defect. Colour code: grey - carbon; blue - nitrogen; red - iron; orange - oxigen}
\label{figure1}
\end{figure}

However, one of the issues concerning the use of CNTs as spintronics devices is the need to incorporate dopants or defects in order to change their electronic transport properties. Closed shell species  do not interact very strongly with the pristine wall of a carbon nanotube.\cite{PhysRevB.67.201401} Furthermore, transition metal atoms are more likely to form clusters when interacting with the pristine wall of the nanotube.~\cite{SolangeDimer} Even linear chains of Fe atoms are more energetically favorable than free standing iron atoms.~\cite{SolangeChain}

One possible path to circumvent this problem is to incorporate doping agents during the growth process. In that context, carbon nanotubes sythetized in a nitrogen-rich atmosphere - the so-called $CN_x$ nanotubes - are potential candidates.\cite{ISI:000245005800008, ISI:000251422000015, ISI:000251355800044, ISI:000173721500002, ISI:000174611600007,arrayscnx} It has already been demonstrated that these nanotubes could be, for example, used as gas sensors for a variety of chemical species.\cite{cnxsensors} 

The most stable defect in these structures is a  pyridine-like defect  consisting of a 4 nitrogen divacancy  (4ND).\cite{ISI:000255524300063,review-alexandre} This defect (shown in fig.~\ref{figure1}a) is formed by two vacancies surrounded by four substitutional nitrogen atoms. We have previously exploited the reactivity of this defect to attach ammonia molecules and study the behavior of the system as a sensor.\cite{ISI:000255524300063} Interestingly, this defect is similar to molecules in the porphyrin class, in particular, to a molecule known as Heme-B (shown in fig.~\ref{figure1}b), which is found, for instance, in hemoglobin and myoglobin. This Heme B molecule has an iron atom bonded to the site with four nitrogens. Thus it is intuitive to assume that an iron atom - and other transition metal (TM) atoms - gets bonded to the 4ND defect of the carbon nanotubes.

The Heme B-like defect has been recently synthesized by Lee et, al.~\cite{PhysRevLett.106.175502} Their stability was studied by means of repeated cyclic voltammograms, and they have not observed significant differences after $10^5$ cycles, which is attributed to the stability of the covalent incorporation of the atoms.  

The use of carbon nanotubes - or any other long one dimensional system - with defects, however, poses an additional problem; the position of the defects cannot be controlled during growth. The result is a device with a large number of randomly positioned defects. Thus, in order to obtain quantitatively meaningful theoretical predictions, one must take into consideration the effects of disorder. 

It is generally assumed that disorder has a detrimental effect. Recently, however, we have shown that one-dimensional Boron-doped graphene nanoribbons can present near-perfect conductance polarization due to disorder.\cite{ISI:000280632500006} Such an effect should not depend on the system under consideration provided there was a difference in transmission probabilities for majority and minority spin cases for a single scatterer. Furthermore, the introduction of a large number of defects with non-zero magnetic moment leads not only to structural disorder, but also to a magnetic one. In fact, the magnetic moments of the impurities might be pointing in random directions. We can then consider that, in the absence of a magnetic field, approximately half of the moments are pointing up and half of them are pointing down. A magnetic field would tend to align the magnetic moments leading to an analogous to the magnetoresistance effect without the need to rely on a multilayered material.

In this work we show that porphyrin-like CNTs can exhibit a nearly-perfect polarization and extremely large disorder induced magnetoresistance (as high as 20000\%) driven by disorder. The disorder was simulated using CNTs containing a large number of magnetic impurities randomly positioned along the tube. For those calculations we used a combination of density functional theory\cite{ISI:A19641557C00018, ISI:A19657000000015} coupled to recursive Green's functions methods\cite{ISI:000230190900016,roche_blase_mplb_2007,ISI:000243195800075,
claudiarochaJPCM,review-alexandre}. 

We have initially performed \textit{ab initio} calculations within Density Functional Theory (DFT)\cite{ISI:A19641557C00018, ISI:A19657000000015} for a segment of a (5,5) $CN_x$ nanotube containing a 4ND defect (figure \ref{figure1}a). As we have said before, the porphyrin molecules can have different TM in the 4ND site, as shown in figure \ref{figure1}b. We, thus, performed calculations using iron, cobalt, manganese and nickel atoms in the middle of the 4ND defect. The final arrangement for the case of iron is presented in figure \ref{figure1}c. As one can see, nine irreducible cells of the pristine system were used to describe the region containing the defect. The computational code used was SIESTA~\cite{ISI:000174901100003, ISI:000252927300009}  which uses a linear combination of atomic orbitals (LCAO) as basis set. In the particular case of this work we have employed a double zeta basis set with polarization orbitals. We used the Generalized Gradient Approximation (GGA) as parametrized by Perdew-Burke-Ernzerhof~\cite{ISI:A1996VP22500044} (PBE) for the exchange correlation functional. Finally, the atomic coordinates have been optimized using a conjugated gradient scheme until the forces on atoms were lower than 0.03 eV/{\AA}.

Furthermore, in order to assess whether the transition metal atoms in adjacent cells are magnetically coupled\cite{ISI:000268429000030}  we simulated a supercell with 18 irreducible cells (twice the initial size) with two Heme-B-like defects; one case with ferromagnetic ordering ($\uparrow\uparrow$), and another with a antiferromagnetic one ($\uparrow\downarrow$). The total energy difference (E$_{\uparrow\uparrow}$-E$_{\uparrow\downarrow}$) is negligible for all transition metal atoms considered in this work, so we infer that there's no magnetic coupling in our system.

For the electronic transport calculations, the system - following the procedure proposed by Carolli \textit{et al}\cite{caroli} - is initially divided in three regions namely, the right and the left electrodes and a central scattering region. The electrodes for our system are taken as semi-infinite repetitions of the pristine carbon nanotube. In the absence of spin-orbit interactions one assumes the two spin fluid approximation, whereby one can calculate the electronic transport properties of the majority and minority spins independently of each other. We then use the Landauer-B\"uttiker~\cite{ISI:A1986E226900027, ISI:A1970F830900022} formula to calculate the transmission coefficients of the system.

\begin{figure}[ht]
\includegraphics[width=0.50\textwidth]{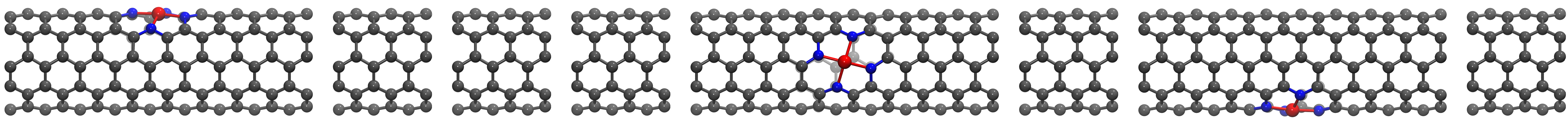}
\caption{A long nanotube built from small pieces randomly distributed and with different rotations, there are also pristine pieces randomly placed.}
\label{long}
\end{figure}

In order to access the electronic transport properties of a more realistic system with hundreds of nanometers we use a  combination of DFT and recursive Green's function methods.~\cite{ISI:000255524300063,ISI:000243195800075,roche_blase_mplb_2007,ISI:000230190900016,review-alexandre} To do so, we split up a long nanotube into small pieces. Each piece is simulated using a separate DFT calculation as already described previously, and the Hamiltonian and overlap opperators respectively $H_S^\sigma$ and $S_S^\sigma$ are stored. For our (5,5) $CN_x$ NT we also have to consider five different rotations for the position of the defect. With those smaller blocks we build up a long nanotube, ranging from 20 to 600 nanometers, by randomly placing the segments with defects together with pristine pieces, as shown in fig.~\ref{long}. One then recursively reduce the system to two renormalized electrodes coupled via an effective scattering potential that contains all the information about the central region.

In the low bias limit the differential spin-dependent conductance can be calculated by the Landauer-Buttiker formula for the current \cite{landauer}
\begin{equation} \label{cond}
g^\sigma=\lim_{V \to 0}\frac{dI^\sigma}{dV}=\frac{e^2}{h}\left.\int T^\sigma(E)\frac{df(E',\mu)}{dE'}\right|_EdE
\end{equation}
where $f(E',\mu)$ is the Fermi distribution function for a given temperature. The total conductance is then given by the sum of the majority and minority conductances, $g=g^\uparrow+g^\downarrow$. 

We are interested in two quantities. Firstly, in order to quantify the spin filtering effect of this device, we calculate the degree of polarization\cite{rocha_jtcn,PhysRevLett.98.196803}
\begin{equation} \label{eq:pol}
P(\%)=\frac{g^{\uparrow}-g^{\downarrow}}{g^{\uparrow}+g^{\downarrow}}\times 100 ~.
\end{equation}

Secondly, as discussed earlier one also needs to take into consideration the relative orientations of the magnetic moments. One can analyze the changes in conductance due to an external magnetic field that tends to align the local magnetization of each impurity. The magnetic field in our calculations is taken into consideration only in the alignment of the magnetic moments, it has no other effect on the electronic structure of our system. Consequently, the value of this disorder induced magnetoresistance (MR) is given by
\begin{equation} \label{eq:gmr}
MR~\left(\%\right)=\frac{g_{100\%}-g_{50\%}}{g_{100\%}}\times 100
\end{equation}
where $g_{100\%}$ corresponds to the total conductance for the case where all of the magnetic moments are pointing along the same direction, and $g_{50\%}$ is the total conductance for the case where there is an equal distribution of positive and negative magnetic moments. 

Finally, in these long and disordered $CN_x$ nanotubes, different defect distributions along the $CN_x$ give different values of conductance. In order to get statistically meaningful values of conductance we have calculated about 200 random arrangements for each concentration and length of the nanotubes.

\begin{figure}[h]
\includegraphics[width=0.50\textwidth]{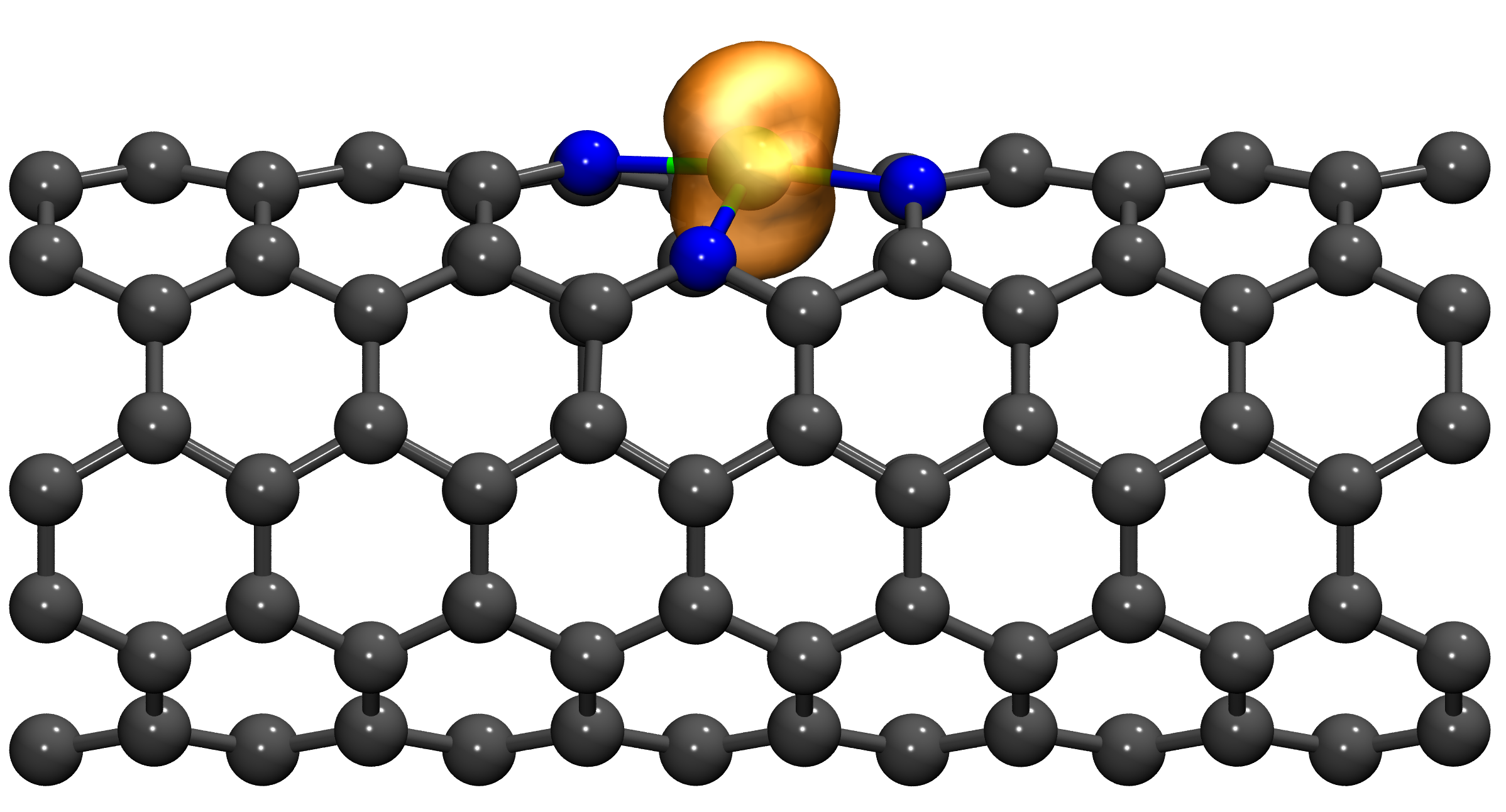}
\caption{Local Magnetization $m\left(\vec{r}\right)=\rho^\uparrow\left(\vec{r}\right)-\rho^\downarrow\left(\vec{r}\right)$ for CN$_x$ nanotube containing an iron atom (Heme-B-like defect). The net magnetic moment of the system is localized in the iron atom.}
\label{rhoupdn}
\end{figure}

Upon placing the different TM atoms in the 4ND, we observe that they strongly bind with a similar final structure in all situations. Table \ref{table1} shows the binding energies and the final magnetization of the system for each one of the TM atoms. As can be seen, the binding energies are relatively high (the reference is the isolated atom infinitelly separated from the nanotube). One also sees a local magnetic moment in all cases, except for the nickel atom. A similar behavior was also observed by Shang, \textit{et al.} .\cite{variostm} In fig.~\ref{rhoupdn} we present the local magnetic moment, $m\left(\vec{r}\right)=\rho^\uparrow\left(\vec{r}\right)-\rho^\downarrow\left(\vec{r}\right)$  of the Heme B-like defect. From this figure we can notice a highly localized magnetic moment in the iron atom. 

\begin{table}[ht]
\caption{Binding energy, magnetization and polarization for different transition metal atoms bonded to the 4ND defect.}
\label{table1}
\centering
\begin{tabular}{ c c c c c } \hline
Atom & Fe & Co & Mn & Ni \\ \hline
Binding Energy (eV) & -6.3 & ~ -4.5 & -4.2 & -4.2 \\
Magnetic Moment ($\mu_B$) & 2 & 3 & 5 & 0 \\
Polarization (\%) & 8.95 & 1.46 & 2.70 & 0.00 \\ \hline
\end{tabular}
\end{table}

\begin{figure}[ht]
\includegraphics[width=0.48\textwidth]{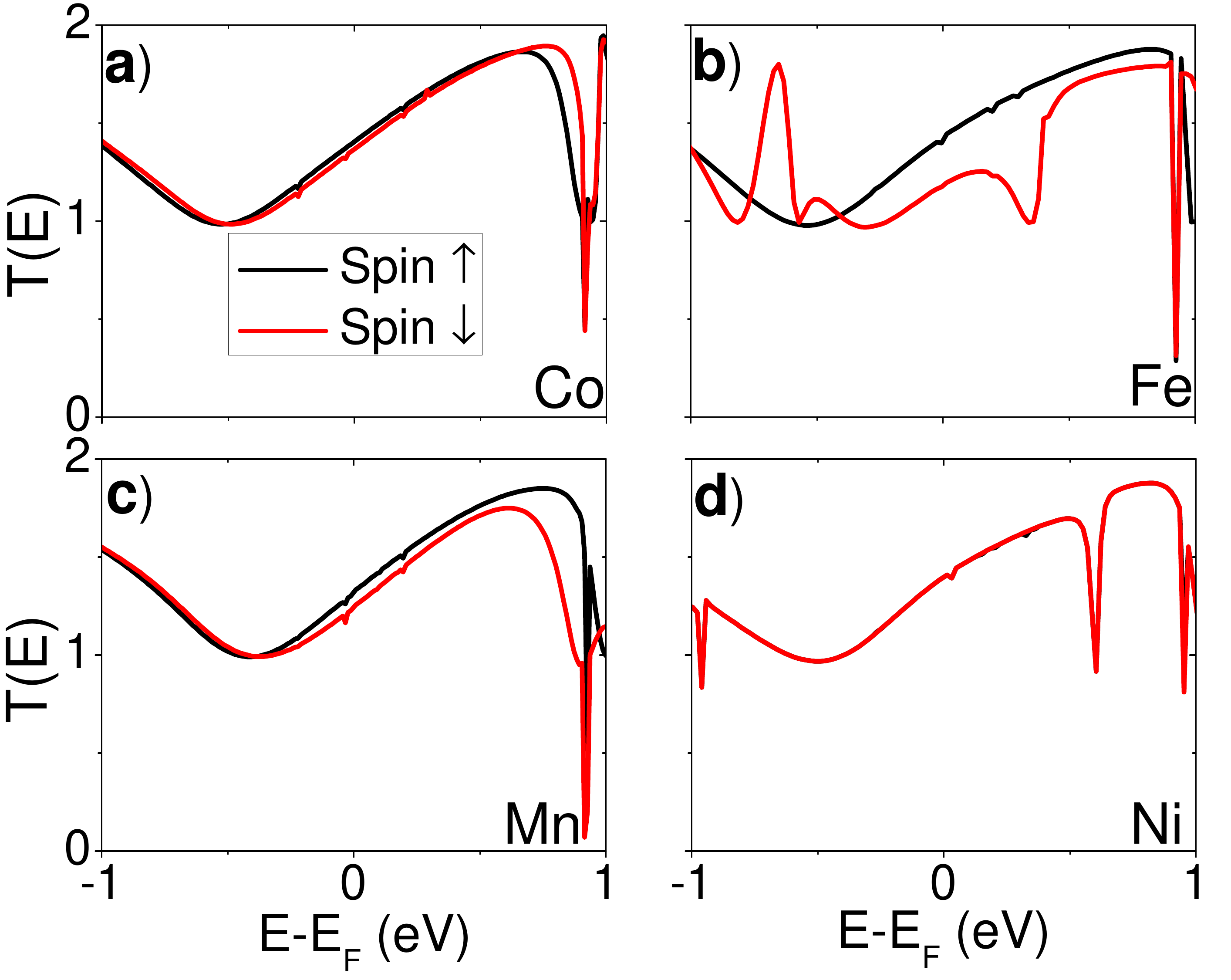}
\caption{Transmission coefficients as function of energy for a single 4ND defect with different transition metal atoms. a) cobalt , b) iron, c) manganese,  and d) nickel.}
\label{transm}
\end{figure}

\begin{figure}[ht]
\includegraphics[width=0.48\textwidth]{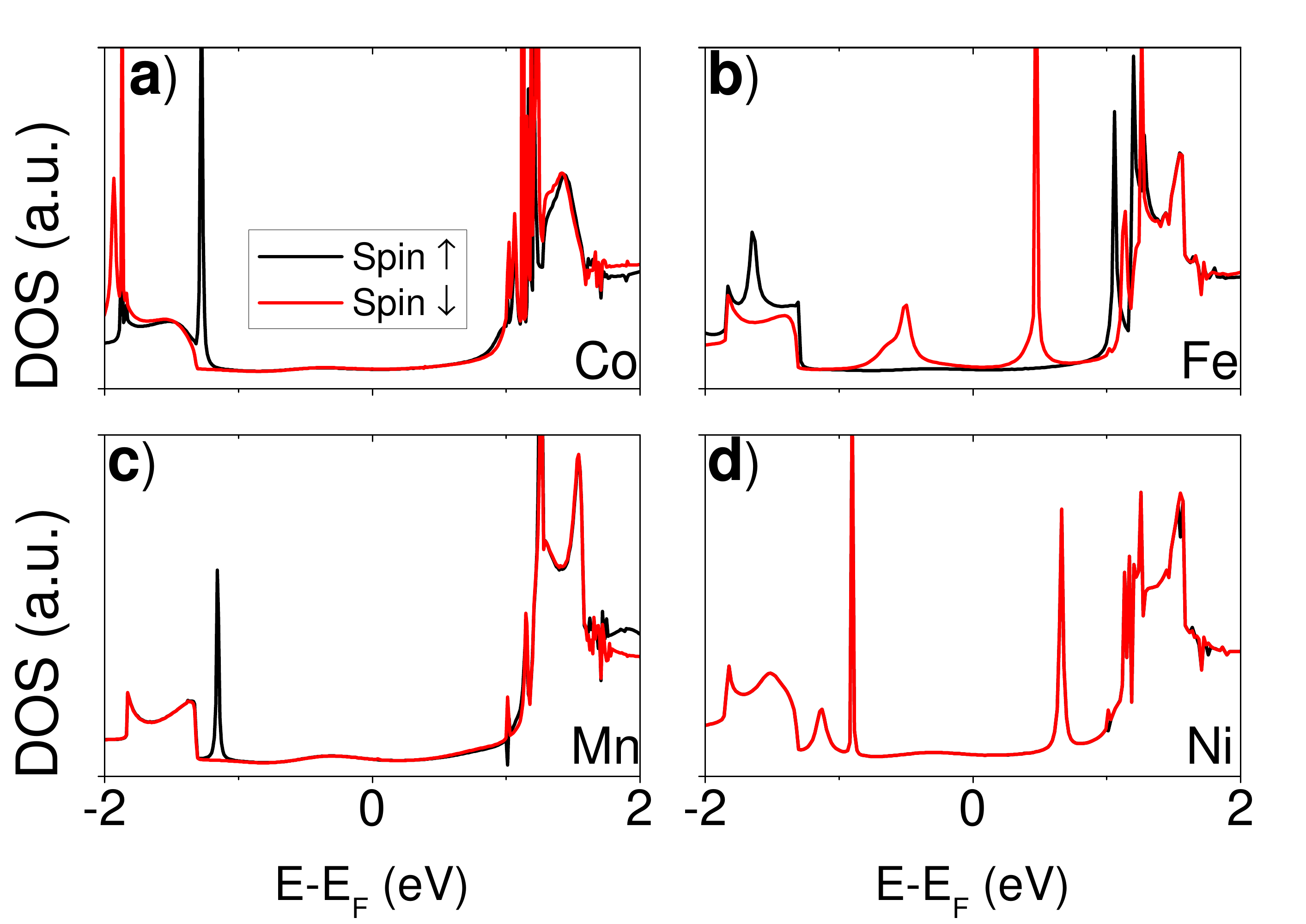}
\caption{Total density of states as a function of energy for a single 4ND defect with different transition metal atoms. a) iron , b) cobalt, c) manganese,  and d) nickel.}
\label{pdos}
\end{figure}

In fig.~\ref{transm} the spin resolved transmission coefficients as a function of energy are plotted for a single 4ND defect with cobalt, iron, manganese and nickel atoms. We can note on fig.~\ref{transm} different transmission probabilities for the up and down spin channels at the Fermi level,\footnote{For the single defect case we use the Fisher-Lee relation for the normalized conductance: $g^\sigma=T^\sigma\left(E_F\right)$.\cite{Fisher:1981uq}} except for nickel, as expected, since nickel shows no local magnetization for this system. For cobalt, iron and manganese the electrons with different spins are not be equally scattered, and this may lead to a spin filter, a system in which a non polarized current enters in one side and after passing through the system the current is spin polarized. We have calculated (with eq.~\ref{eq:pol}) the polarizations shown in table~\ref{table1} using the conductances obtained for a low temperature calculation. In all cases there was only a single TM defect and one can clearly see that the polarization is always bellow 10\%. Although the magnetic moments are generally high - except for nickel - the localized $d$ states are positioned either above or bellow  the Fermi level. This can seen from fig.~\ref{pdos} where the total density of states are shown. In the particular case of iron (figure \ref{pdos}b) the minority spin states are closer to the Fermi level which consequently lead to a larger polarization.\footnote{He have also performed calculations using a self-interaction corrected LDA,\cite{ISI:000232229800045,ISI:000243895600018} but we have observed no significant changes to the polarization.} Still, this leads to values of the polarization which are not very high for any of these systems to behave as spin filters.

\begin{figure}[ht]
\includegraphics[width=0.48\textwidth]{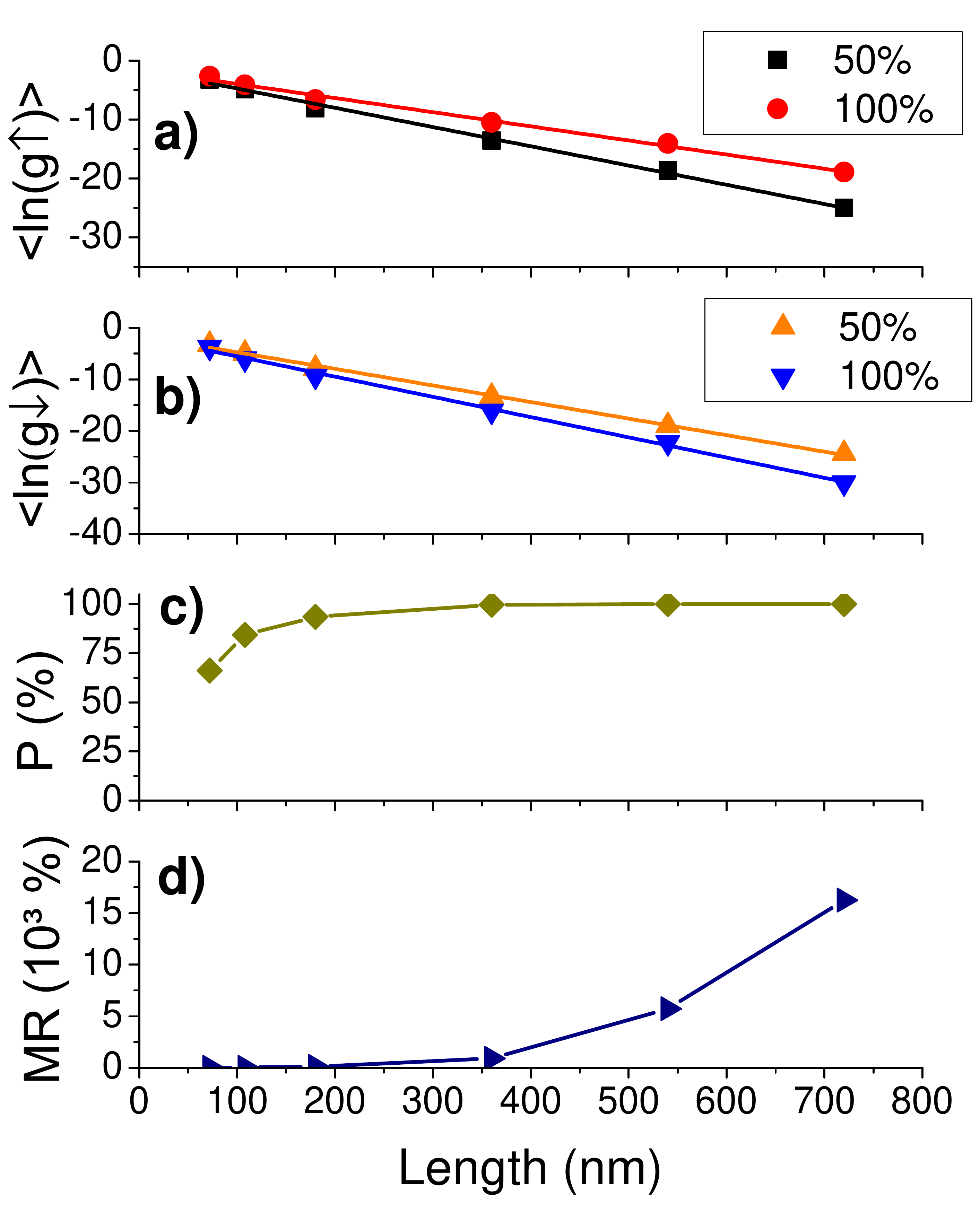}
\caption{Average logarithm for a) majority and b) minority spin conductances as a  function of length. c) Polarization for the 100 \% case, and d) average magnetoresistance. All graphs shown correspond to a defect concentration of 0.65\% and temperature of 3K.}
\label{condpol100}
\end{figure}

For the long disordered $CN_x$ we have chosen the Heme-B defects (Fe), since they show the highest polarization for a single defect. In all cases considered here we have taken a defect concentration of 0.65\%  per mass.
In figure \ref{condpol100}(a-b) we show the spin-resolved logarithm of the conductance averaged over more then 200 random realizations as a function of device length. As we are in the Anderson localization regime the conductance is supposed to vary exponentially as function of length, following the well know exponential relation
\begin{equation}
g^{\sigma}\propto e^{-\frac{L}{\xi^\sigma}} ~,
\end{equation}
where $\xi^\sigma$ is the spin-dependent localization length. This can be clearly seen for both the cases of majority and minority spin channels. The localization lengths obtained by the linear fit are shown in table \ref{con65_3k_100}.

As one can clearly see from figure \ref{condpol100}c the conductance is completely spin polarized for the case where all the magnetic moments are pointing in the same direction and the nanotube is longer than 200 nm. For the case where half the magnetic moments are pointing up and half are pointing down the average magnetization is, as expected, zero (and therefore not shown in the graph). As shown in a previous work,\cite{ISI:000280632500006} this disorder-driven polarization effect is independent of the type of system being studied provided there are different scattering probabilities for majority and minority spin channels in the single impurity case. For the 100\% case, since $\xi^\uparrow$ is larger than $\xi^\downarrow$, the spin down conductance will decay much faster than the spin up conductance, following 
\begin{equation}
g^\uparrow\propto \left( g^\downarrow\right)^n,
\end{equation}
where $n=\xi^\uparrow/\xi^\downarrow$. For the case shown here $n=1.642$, whereas for the single impurity case, $R^\downarrow/R^\uparrow=1.634$. These two results are very close, corroborating our previous assumption.\cite{ISI:000280632500006}

\begin{figure}[ht]
\includegraphics[width=0.48\textwidth]{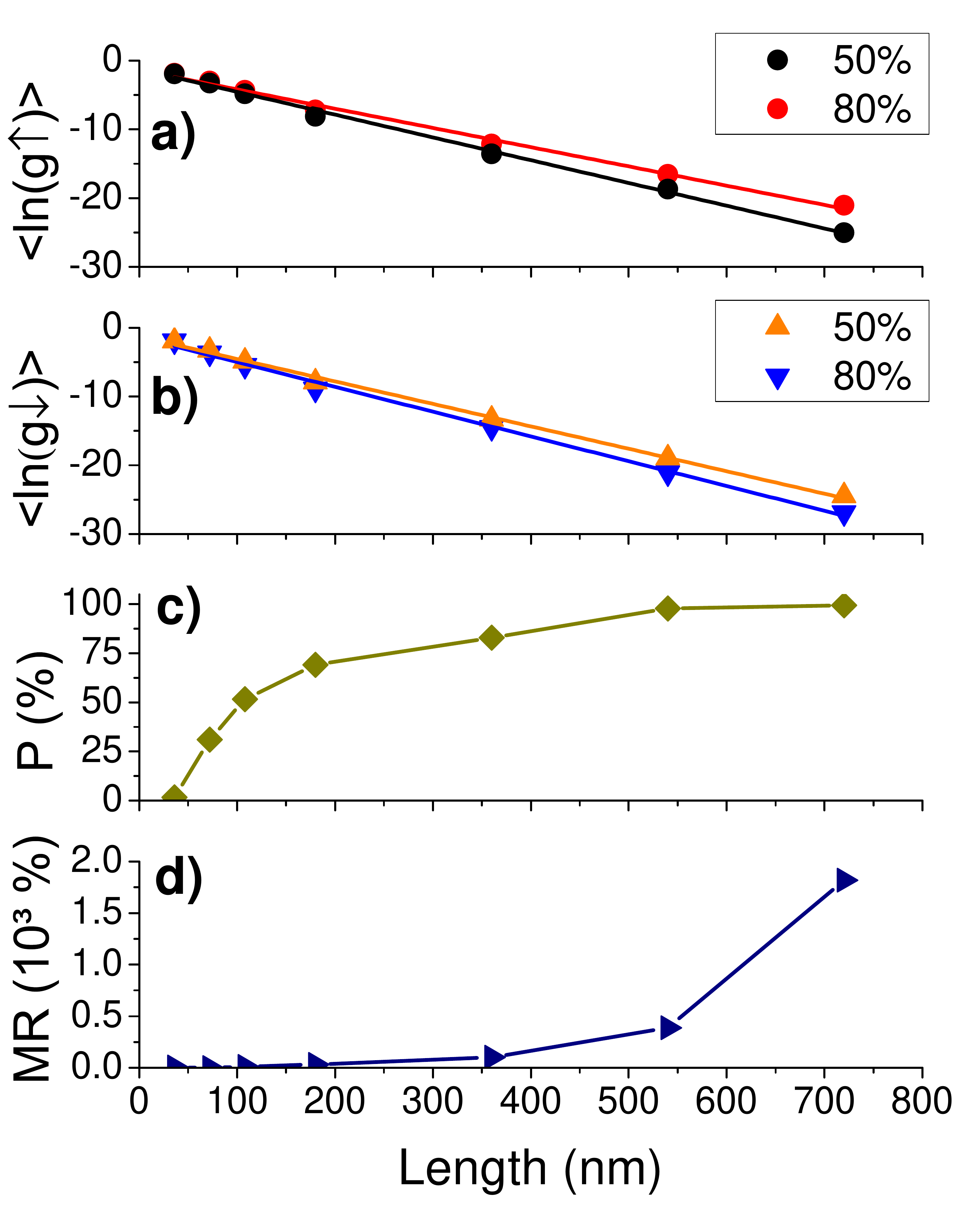}
\caption{Average logarithm for a) majority and b) minority spin conductances as a  function of length. c) Polarization for the 80 \% case, and d) average magnetoresistance. All graphs shown correspond to a defect concentration of 0.65\% and temperature of 3K.}
\label{condpol2}
\end{figure}

In fig.~\ref{condpol100}d we present the values for the magnetoresistance (calculated with eq~\ref{eq:gmr} as a function of length. The calculated MRs in this case reached values up to 20000~\%. As in the case of the polarization, the presence of randomly distributed scatterers lead to an enhancement of the MR effect up to extremely large values.

\begin{table}[ht]
\caption{Values of the spin-dependent localization lengths $\xi^\sigma$ for 0.65\% defect concentration and temperature of 3K.}\label{con65_3k_100}
\centering
\begin{tabular}{cccc}
\hline
 & \multicolumn{2}{c}{ Localization length (\r{A})} \\
 & 100\% & 80\% & 50\%  \\ \hline
Spin up & 417 &  357 & 302 \\
Spin down & 254 & 277 & 306 \\
\hline
\end{tabular}
\label{lengths100}
\end{table}

In order to address the effect of a magnetic field applied to the system and the possibility that not all the magnetic moments are completely aligned we also considered that only 80\% of the defects are magnetically aligned. The results are shown in figure \ref{condpol2}, and the linear fits for the localization lengths also are presented in table \ref{lengths100}. We can note a decrease (increase) in localization length for majority (minority) spin compared to the 100 \% case. This is to be expected since one is moving towards higher magnetic disorder, {\it i.e.} the 50 \% case. From fig.~\ref{condpol2}c we can see a polarization near 100\% for 700nm long $CN_x$ nanotubes. Most importantly, the magnetoresistance (Fig. \ref{condpol2}d) presents values which are one order of magnitude lower than the fully aligned arrangement, but it is still in the 1000 \% range. Thus, even in the case where not all spins are aligned, there is still an extremely large disorder induced magnetoresistance. Thus, this disorder-driven GMR effect is extremely robust toward fluctuations in the alignment of the magnetic moments. 

We have used an ideal paramagnet model to estimate the needed magnetic field to obtain a 80\% magnetization at 300K and 3K. Unfortunately the needed magnetic field at ambient temperature is about 200 T, making it impracticable for ambient temperature devices. For a temperature of 3K the needed field will be about 2T for an 80\% magnetization and about 5T for a 95\% magnetization, so it's possible to observe the predicted effects in this paper in low temperature experiments.

In summary we have observed that transition metal atoms bind strongly to nitrogen defects in CN$_x$ carbon nanotubes in a fashion similar to Heme B molecules. The end result is a scattering site with a localized magnetic moment on the transition metal atom that leads to a small conductance polarization in the case of a single impurity. For a large number of such impurities randomly distributed along carbon nanotubes a few hundred nanometers long, it leads to near perfect polarization and a large magnetoresistance (up to 20000\%). An interesting feature of the system proposed here is the fact that they do not need polarized electrodes as it is usually the case. We estimate that, at low temperature, this effect could be measured experimentally. 
 
\section{Acknowledgements}

The authors would like to acknowledge FAPESP, CAPES and CNPq for financial support. The calculations were carried out at CCE-USP, Center for High Performance Computing at UFABC and CENAPAD/SP.

\end{document}